\documentclass{article}
\usepackage{epsfig}

\usepackage{latexsym}

\newcommand{\ap}[3]{{\sl Ann.~Phys.} {\bf #1} (19#2) #3}

\newcommand{\nc}[3]{{\sl Nuovo Cimento} {\bf #1} (19#2) #3}
\newcommand{\np}[3]{{\sl Nucl. Phys.} {\bf #1} (19#2)~#3}
\newcommand{\pl}[3]{{\sl Phys. Lett.} {\bf #1} (19#2) #3}
\newcommand{\pr}[3]{{\sl Phys. Rev.} {\bf #1} (19#2) #3}

\newcommand{\vj}[4]{{\sl #1~}{\bf #2} (19#3) #4}

\def\be{\begin{equation}}
\def\ee{\end{equation}}
\def\bea{\begin{eqnarray}}
\def\eea{\end{eqnarray}}

\newcommand{\bfi}[1]{\begin{figure}[#1]}
\newcommand{\efi}{\end{figure}}
\newcommand{\bpi}[2]{\begin{picture}(#1,#2)}
\newcommand{\epi}{\end{picture}}

\def\ie{{\it i.e.\/}}

\newcommand{\prop}{\Delta}
\newcommand{\propm}{{\Delta}_m}

\renewcommand{\d}{\partial}

\newcommand{\OO}{{\mathcal O}}

\begin{document}


{\hfill \parbox{6cm}{\begin{center} 
        UG-FT-95/98  \\
        UAB-FT-463   \\
        hep-ph/9901291 \\
        January 1999                   
\end{center}}} 
              
\vspace*{1cm}                               
\begin{center} 
\large{\bf Constrained Differential Renormalization \\ }
\large{\bf and Dimensional Reduction}\footnote{Talk presented at
the IVth International Symposium on Radiative Corrections, Barcelona, Spain,
September 1998.} 
\vskip .3truein  
{\large 
F. del Aguila}  
\end{center}  
\begin{center}
{\it Depto. de F\'{\i}sica Te\'orica y del Cosmos,  
 Universidad de Granada \\
 18071 Granada, Spain} 
\end{center}
\vspace{.1cm} 
\begin{center}
{\large 
M. P\'erez-Victoria\footnote{On leave of absence from Depto.\ de 
F\'{\i}sica Te\'orica y del Cosmos,  
 Universidad de Granada,
 18071 Granada, Spain}}  
\end{center}  
\begin{center}
{\it Grup de F{{\'\i}}sica Te{\`o}rica and Institut de F{{\'\i}}sica
        d'Altes Energies (IFAE), \\
        Universitat Aut{\`o}noma de Barcelona \\
        08193 Bellaterra, Barcelona, Spain}
\end{center}
\vspace{.8cm} 
 
\centerline{\bf Abstract} 
\medskip 
We describe the equivalence at one loop between
constrained differential renormalization and regularization by dimensional
reduction in the MS scheme. To illustrate it, we reexamine
the calculation of supergravity corrections to $(g-2)_l$.

\vspace*{1cm}

\section{Introduction}
The most popular regularization method for perturbative calculations in gauge
theories is dimensional regularization~\cite{dimreg}. In conjunction with the
minimal substraction (MS) scheme, 
it leads to renormalized Green functions
satisfying the Slavnov-Taylor identities. It has problems, however, 
in supersymmetric gauge theories, because invariance under supersymmetry
transformations depends on the specific dimensionality of the objects involved.
Hence, in general, dimensional regularization does not preserve 
supersymmetry. To improve the situation, Siegel proposed
a modified version of the method~\cite{dimred}, called regularization 
by dimensional reduction (or dimensional reduction, for short). 
It treats integral momenta (or space-time points)  as $d$-dimensional
vectors but takes all fields to be four-dimensional tensors or spinors. The
relation between the four-dimensional and $d$-dimensional spaces is given by
dimensional reduction from 4 to $d$ dimensions, \ie, the 4-dimensional space is
decomposed into a direct sum of $d$- and (4-$d$)-dimensional subspaces in the
following sense (we always work in Euclidean space)~\cite{Avdeev83}: 
the 4-dimensional metric-tensor 
$\delta_{\mu\nu}$ (satisfying the properties $\delta_{\mu\nu} 
\delta_{\nu\rho}=\delta_{\mu\rho}$ and $\delta_{\mu\mu}=4$) and the 
$d-$dimensional one $\tilde{\delta}_{\mu\nu}$ (satisfying 
$\tilde{\delta}_{\mu\nu} \tilde{\delta}_{\nu\rho}=\tilde{\delta}_{\mu\rho}$ 
and $\tilde{\delta}_{\mu\mu}=d$) are related by
\be
        \delta_{\mu\nu}\tilde{\delta}_{\nu\rho} = \tilde{\delta}_{\mu\rho}.
        \label{dimredeq}
\ee
Although this method is known to be inconsistent~\cite{Siegel2}, the
inconsistencies are in many cases under control and dimensional reduction is
actually the preferred regularization method for explicit calculations in
supersymmetric theories\footnote{In Ref.~\cite{Avdeev} some modifications were
proposed which make the scheme consistent at the price of breaking
supersymmetry (at higher orders).}

Differential renormalization is a position-space method that performs 
regularization and substraction in one step by substituting ill-defined
expressions by derivatives of well-defined ones~\cite{FJL}.
Recently, a new version aiming to preserve gauge invariance~\cite{QCD} 
and supersymmetry~\cite{g2,Barna}
has been developed~\cite{CDR,techniques} and automatized~\cite{auto} 
at the one-loop 
level. This version, called constrained differential renormalization (CDR), 
is based on a set of rules that determine the renormalization of the 
Green functions. 
In Ref.~\cite{auto}, T. Hahn and one of the authors (M.P.V.) argued and 
explicitly checked that, to one loop, 
CDR and dimensional reduction in the MS scheme 
render the same results, up to a redefinition
of the renormalization scales. Our purpose here is to discuss this
in greater detail. In Section~2 we do it in position space. Its
counterpart in momentum space is briefly discussed in Section~3.  
In Section~4, we illustrate the equivalence by comparing the calculation 
of the anomalous magnetic moment of a charged lepton in supergravity when 
CDR and dimensional reduction are employed. 
Finally, Section~5 is devoted to conclusions.

\section{CDR and dimensional reduction in position space}
CDR, as the usual differential renormalization method, is naturally formulated
in position space. CDR renormalizes each Feynman graph by reducing it
to a linear combination of basic funcions (and their
derivatives) which are then replaced by their
renormalized expressions. The renormalization of the basic functions is
fully determined by four rules to be described below, which are significant
for the fulfilment of Ward identities. A generic (one-loop) 
basic function is a string of propagators, with a differential operator $\OO$ 
acting on the last one. $\OO$ is either the identity or a ``product"
of space-time derivatives. Basic functions with differential
operators with contracted or uncontracted indices are considered
independent, because it turns out that
contraction of Lorentz indices does not commute with CDR. For this reason,
to decompose a Feynman graph into basic functions one must simplify all the
(Dirac) algebra and contract all Lorentz indices. Notice that reducing the
renormalization of a Feynman graph to the renormalization of the basic functions
is equivalent to linearity and compatibility of CDR
with the Leibniz rule for the derivative of a product (which is used in the
decomposition).

The renormalization of the basic functions is determined by four
rules~\cite{CDR}:
\begin{enumerate}
  \item {\em Differential reduction}: 
  singular expressions are substituted by derivatives
  of regular ones. 
  \label{R1} We distinguish two cases:
     \begin{enumerate}
       \item Functions with singular behaviour worse than 
        logarithmic ($ \sim x^{-4}$) are reduced to derivatives 
        of logarithmically
        singular functions without introducing extra dimensionful
        constants. \label{R1a}
       \item Logarithmically singular functions are written as
        derivatives of regular functions.
        This requires introducing an arbitrary dimensionful constant.
        \label{R1b}
      \end{enumerate}   
  \item {\em Formal integration by parts}: derivatives act
   formally by parts on test functions. \label{R2} In particular,
\be
  [\d F]^R = \d F^R \, ,
\ee
   where $F$ is an arbitrary function and $R$ stands
   for renormalized.
  \item {\em Delta function renormalization rule}: \label{R3}
\be
  [F(x,x_1,...,x_n) \delta(x-y)]^R =  [F(x,x_1,...,x_n)]^R
  \delta(x-y)\, .
\ee 
  \item The general validity of the {\em propagator equation}:
   \label{R4}
\be 
  \left[ F(x,x_1,...,x_n) (\Box^x - m^2) \propm(x) \right]^R = 
  \left[ F(x,x_1,...,x_n) (- \delta(x)) \right]^R\, ,
  \label{masspropeq}
\ee
   where $\propm(x) = \frac{1}{4\pi^2} \frac{m K_1(mx)}{x}$ and
   $K_1$ is a modified Bessel function.
\end{enumerate}
Rule~\ref{R1}  
reduces the ``degree of singularity'',
connecting singular and regular expressions. 
Rule~\ref{R2} is essential to make
sense of rule~\ref{R1}, for 
otherwise the right-hand-side of it would not be a well-defined
distribution. These two rules are the essential prescriptions of
the method of differential renormalization. 
Forbidding the introduction of dimensionful scales outside logarithms, we
completely fix the scheme\footnote{This prescription simplifies calculations
and the renormalization group equation. Nevertheless, in all cases we have
studied (scalar and spinor QED and QCD) the inclusion of dimensionful
constants outside logarithms does not spoil gauge invariance, as long as the
other rules are respected.}. 
Note that the last three rules are valid mathematical identities among 
tempered distributions when applied to a well-behaved enough
function $F$. The rules formally extend their range of 
applicability to arbitrary functions.

Rule~\ref{R1} specifies the renormalization of any one-loop expression
up to arbitrary local terms.
The other rules lead to a system of algebraic equations for these local 
terms~\cite{techniques}.
It turns out that a solution exists, 
and this solution is unique once an initial condition is given (apart from the
requirement in rule~\ref{R1a} of not introducing extra dimensionful constants,
which is also an initial condition):
\be
  \left[\prop_0(x)^2 \right]^R = \left[\left(\frac{1}{4\pi^2x^2}\right)^2 
  \right]^R = 
  - \frac{1}{(4\pi^2)^2} \frac{1}{4}  \Box \frac{\log x^2 M^2}{x^2}\, .
\label{DRidentity}
\ee 
This is the most general realization of rule~\ref{R1b} for $\prop_0(x)^2$,
and introduces the unique dimensionful constant of the whole process, $M$, 
which has dimensions of mass and plays the role of renormalization group scale.

The decomposition of Feynman graphs into basic functions can be performed
in both dimensional regularization and dimensional reduction in exactly the 
same way as we have described for CDR. Although in the dimensional methods this
prescription is not necessary (for in $d$ dimensions everything is
well-defined), we shall assume that all Lorentz indices have
been contracted before identifying the basic functions. These contain only
d-dimensional objects both in dimensional regularization and in dimensional
reduction. Indeed, although in the latter constractions with the
4-dimensional metric tensor are performed, Eq.~\ref{dimredeq} projects them 
into the $d$-dimensional subspace. Hence, the regulated basic functions
are identical in these two methods.
On the other hand, expressions dimensionally regulated satisfy
rules \ref{R2} to~\ref{R4} because they are well-defined 
distributions. They also satisfy rule~\ref{R1a} for the same reason (what 
agrees with the scaling property
of $d$-dimensional integrals, which forbids the appearance of new 
dimensionful constants). A renormalization scale $\mu$ is 
introduced to keep the coupling constant with a fixed
dimension and appears only inside logarithms. 
Rule~\ref{R1b} is never needed because the (formal) degree of
divergence is non-integer in the dimensional methods. 
Instead, the use of rule~\ref{R1a} in expressions which diverge logarithmically
when $\epsilon=\frac{4-d}{2}\rightarrow 0$, gives rise to poles 
of the form $\frac{1}{\epsilon}$. 
In particular, the regularized value of $\prop_0(x)^2$ is~\cite{Nuria}
\be
\mu^{2\epsilon} \frac{\Gamma^2(\frac{d}{2}-1)}{4^2\pi^d} x^{4-2d} =
\frac{1}{(4\pi^2)^2} \left[\pi^2 \frac{1}{\epsilon}\delta^{(d)}(x)
-\frac{1}{4}\Box\frac{\log (x^2\mu^2\pi\gamma_E e^2)}{x^2} \right]
+ O(\epsilon) \, ,
\label{dimregidentity}
\ee
where we have included the global factor $\mu^{2\epsilon}$ to have a 
dimensionless argument in the logarithm, expanded in $\epsilon$ and used the
$d$-dimensional equalities 
\be
        x^{-p}=\frac{\Box x^{-p+2}}{(-p+2)(d-p)}
\ee
and
\be
        \Box \left[\frac{\Gamma(\frac{d}{2}-1)}{4\pi^{d/2}} x^{2-d}\right] =
        -\delta^{(d)}(x) \, , \label{dprop}
\ee
to rewrite it. $\gamma_E=1.781\ldots$ is Euler's constant.

Now, since the dimensionally regularized basic
functions satisfy the CDR rules, they must also be a solution 
to the set of algebraic 
equations discussed above, but with the initial condition given by 
Eq.~\ref{dimregidentity}. This is true for each order of
the Laurent series in $\epsilon$. Therefore, substracting the
$\frac{1}{\epsilon}$ poles, which always multiply a local term,  
and taking the limit $\epsilon \rightarrow 0$ 
(\ie, using the MS scheme) one obtains renormalized basic functions that
are $\epsilon$-independent solutions of the equations. In particular, 
the renormalization of $\prop_0(x)^2$ is given by Eq.~\ref{DRidentity} 
with 
\be
M^2=\mu^2 \pi \gamma_E e^2 \, .
\label{relation}
\ee
Once the initial condition is completely fixed, the
solution to the set of equations is unique, so it must be the same in CDR and
in dimensional regularization or dimensional reduction.
Summarizing, {\bf the renormalized basic functions in dimensional 
regularization, dimensional reduction and CDR 
are identical if the MS scheme is used for the former methods 
and Eq.~\ref{relation} holds}.

This does not mean that the renormalized Feynman diagrams are the same in the
three methods, because in the dimensional ones the substraction must be
performed after multiplying by the coefficients 
outside the basic functions. Then, 
if these coefficients contain $O(\epsilon)$ pieces, 
the structure of the Laurent 
series can be spoiled. Extra local $O(\epsilon^0)$ terms are picked up from
the $\frac{1}{\epsilon}$ poles of the basic functions, and the final result 
does not, in general, coincide with the CDR one.
However, this does not occur in the case of 
dimensional reduction because, if the
decomposition of the diagram has been performed as in CDR, there are no
$O(\epsilon)$ pieces out of the basic functions. The reason is that all the
coefficients are 4-dimensional and are never projected into $d$ dimensions
since all contractions with $d$-dimensional objects were already performed
and included in the definition of the basic functions. In other words, with this
decomposition all external indices are 4-dimensional (and all internal ones are
$d$-dimensional). Therefore, {\bf renormalized Feynman diagrams in CDR and in 
dimensional reduction with MS coincide, if
Eq.~\ref{relation} holds}. This is not true in dimensional regularization
because the dimension $d$ can appear explicitly outside the basic functions, for
everything is considered $d$-dimensional. For example,
$\tilde{\delta}_{\mu\mu}=d$ can appear out of the basic functions.
Dimensional regularization only coincides with dimensional reduction and CDR at
the level of basic functions.

\section{CDR and dimensional reduction in momentum space}
Obviously, if CDR and dimensional reduction give the same renormalized
amplitudes in position space, they do too in momentum space, 
because the Fourier transforms 
of well-defined distributions are uniquely determined.
The decomposition into basic functions is performed exactly as in 
position space, the basic functions corresponding now to (tensor) basic
integrals of a set of internal momenta times a product of propagators. 
This can be done because the linearity of CDR in 
position space (together with rule~\ref{R2}) implies 
linearity in momentum space. Then, one just has to substitute divergent
basic integrals by the Fourier transforms of the corresponding renormalized
expressions in position space. In the case of dimensional reduction this
is the same as doing the full calculation in momentum space. The Fourier
transform of the initial condition Eq.~\ref{DRidentity} is
\be
\left[ \int \frac{d^4 k}{(2\pi)^4} \frac{1}{k^2(k-p)^2}  \right]^R
= \frac{1}{16\pi^2} \log \frac{\bar{M}^2}{p^2} \, ,
\ee
where $\bar{M}=2M/\gamma_E$. The relation between this $\bar{M}$ and
$\mu$ is given by
\be
\bar{M}^2 = \frac{\mu^2 4\pi}{\gamma_E} e^2 = \bar{\mu}^2 e^2 \, ,
\label{momrelation}
\ee
where $\bar{\mu}$ is the renormalization scale of the $\overline{\mbox{MS}}$
scheme. This is the relation found in Ref.~\cite{auto}.

\section{A physical example: $(g-2)_l$ in supergravity}
The calculation of the anomalous magnetic moment of a charged lepton, 
\mbox{$(g-2)_l$},
in supergravity is a convenient place to compare CDR with dimensional
regularization and dimensional reduction. First, $(g-2)_l$ is an observable;
second, it is power counting divergent (and hence regularization dependent); 
and third, supersymmetry requires it to vanish~\cite{Ferrara}. 
The diagrams giving $e\kappa^2$
corrections, where $e$ is the electric charge and $\kappa^2=8\pi G_N$, with
$G_N$ Newton's constant, are depicted in Fig.~1.
\begin{figure}
\centering
\mbox{
\epsfig{file=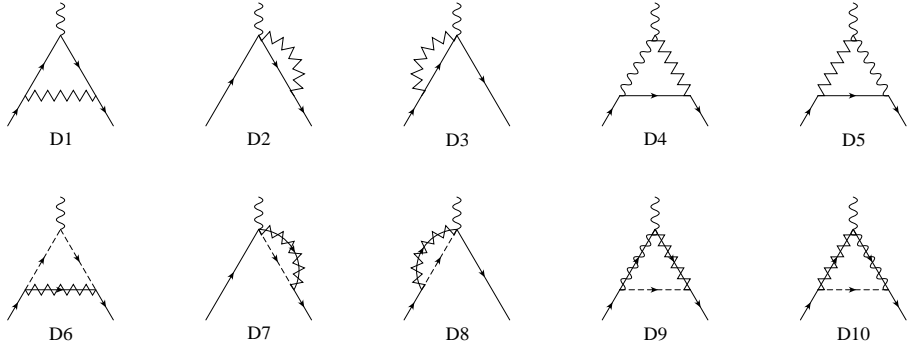 ,width=12cm   }
}

\caption{Diagrams of order $e \kappa^2 $ contributing to $(g-2)_l$ 
in SUGRA. A graviton is exchanged in diagrams D1-D5 and a gravitino
in D6-D10.}   
\end{figure}
This calculation has been discussed in detail several
times~\cite{Mendez,Bellucci,Wilcox,Grisaru,g2,Barna}.
Table~1 gathers the different contributions in dimensional regularization,
dimensional reduction and CDR.
\begin{table}[h,t]

\vspace{0.4cm}

\begin{center}

\footnotesize

\begin{tabular}{|l| l |l |l|}
\hline  
& \multicolumn{1}{|c|}{Dimensional} & \multicolumn{1}{|c|}{Dimensional} & \\
\multicolumn{1}{|c|}{Diagram} 
& \multicolumn{1}{|c|}{regularization} &
\multicolumn{1}{|c|}{reduction} &
\multicolumn{1}{|c|}{CDR} \\  
        \hline  & & & \\
D1 & $-\frac{1}{6} (\frac{1}{\epsilon}+
        \log\frac{\bar{\mu}^2}{m^2})-\frac{61}{36}$
        & $-\frac{1}{6}(\frac{1}{\epsilon}+
        \log\frac{\bar{\mu}^2}{m^2})-\frac{29}{18}$  
        & $-\frac{1}{6} \log\frac{\bar{M}^2}{m^2}
        -\frac{23}{18}$  \\ 
D2+D3 & $-\frac{11}{6} (\frac{1}{\epsilon}+
        \log\frac{\bar{\mu}^2}{m^2})-\frac{32}{9}$
        & $-\frac{11}{6} (\frac{1}{\epsilon}+
        \log\frac{\bar{\mu}^2}{m^2})-\frac{35}{9}$ 
        & $-\frac{11}{6} \log
        \frac{\bar{M}^2}{m^2}-\frac{2}{9}$  \\
D4+D5   & $2 (\frac{1}{\epsilon}+
        \log\frac{\bar{\mu}^2}{m^2})+7$
        & $2 (\frac{1}{\epsilon}+
        \log\frac{\bar{\mu}^2}{m^2})+6$ 
        & $   2  \log \frac{\bar{M}^2}{m^2} +2$   \\ 
        & & &    \\
  Graviton    & & & \\     
  (D1+D2+D3   &$7/4$              &1/2      &   1/2   \\
  +D4+D5)   & & & \\ 
        & & &\\ 
        \hline & & & \\
  D6        &$-\frac{4}{3}  (\frac{1}{\epsilon}+
        \log\frac{\bar{\mu}^2}{m^2})-\frac{55}{18}$
        &$-\frac{4}{3}  (\frac{1}{\epsilon}+
        \log\frac{\bar{\mu}^2}{m^2})-\frac{37}{18}$ 
        &  $-\frac{4}{3} \log\frac{\bar{M}^2}{m^2}+
        \frac{11}{18}$   \\ 
  D7+D8     &$-\frac{2}{3}  (\frac{1}{\epsilon}+
        \log\frac{\bar{\mu}^2}{m^2})-\frac{13}{9}$
           &$-\frac{2}{3} (\frac{1}{\epsilon}+
        \log\frac{\bar{\mu}^2}{m^2})-\frac{4}{9}$ 
           &   $-\frac{2}{3} \log\frac{\bar{M}^2}{m^2}+
           \frac{8}{9}$  \\
  D9+D10   &$2 (\frac{1}{\epsilon}+
        \log\frac{\bar{\mu}^2}{m^2})+4$
           &$2 (\frac{1}{\epsilon}+
        \log\frac{\bar{\mu}^2}{m^2})+2$ 
           &$   2  \log \frac{\bar{M}^2}{m^2} -2$   \\
        & & & \\
  Gravitino    & & &   \\ 
  (D6+D7+D8    &$-1/2$             &$-1/2$          &   $-1/2$
  \\
  +D9+D10)   &  & & \\ 
        & & &\\  \hline 
  TOTAL      &  &  & \\
  (Graviton   & 5/4 &0 &   0  \\ 
  +Gravitino)   & & & \\
 \hline
\end{tabular} 

\end{center}
\caption{Contributions of the diagrams in Fig.~1 
to $\left( \frac{g-2}{2} \right)_l$ in units of 
$\frac{G_N m^2}{\pi}$, obtained with dimensional regularization, dimensional
reduction and CDR.}
\end{table}
Although the contribution of each diagram diverges, the sum of all diagrams
where a graviton (D1-D5) is interchanged is finite~\cite{Berends} 
(in all methods), as is the sum of those with a gravitino interchange 
(D6-D10), and hence the total sum. 
Whereas dimensional regularization breaks supersymmetry and gives a 
non-zero result~\cite{Mendez,Bellucci}, a vanishing correction is obtained 
both in dimensional reduction~\cite{Mendez,Bellucci} and in CDR~\cite{g2,Barna}.
We see that CDR and dimensional reduction in MS do give the same 
results for each diagram if the renormalization scales are related by 
Eq.~\ref{momrelation}. The total graviton (gravitino)
contribution being finite, it is identical in both methods. (In
Ref.~\cite{Barna} the scale-independent parts of the CDR result have errors due
to the omission of a local term in one basic function, but the total graviton and
gravitino contributions are correct because that local term cancels in the
sums.) 

\section{Conclusions}
We have discussed the one-loop 
equivalence of CDR and dimensional reduction in the
MS ($\overline{\mbox{MS}}$) scheme. 
The result also applies in the presence of
anomalies, for both methods can be used with the same computing rules in
position (momentum) space: Feynman
diagrams are decomposed completely into basic functions (integrals), 
doing all the algebra in 4 dimensions, and then the singular (divergent) 
expressions are replaced by the renormalized ones. In the two methods, chiral
anomalies appear as ambiguities in the writing of the external tensors: it 
is possible to add pieces which vanish in 4 dimensions but change the 
decomposition into basic functions (integrals), and this can affect the final
result due to the non-commutation of renormalization with contraction
of Lorentz indices. In dimensional reduction this can be also understood as the
fact that these pieces are projected into $d$ dimensions, where they do not
vanish any longer. In Ref.~\cite{Rheinsberg} we showed how 
the right ABJ anomaly~\cite{ABJ} was recovered in the context of CDR 
and checked that a democratic treatment of the
traces located all the anomaly in the chiral current. 
Exactly the same applies to dimensional reduction.

CDR has been only developed at the one-loop level, but an extension of the
method to higher orders, 
based on the same rules ~\ref{R1}--\ref{R4} or their extension, is in
principle possible. It does not follow from our discussion that such a
method should be equivalent to dimensional reduction. On the one hand,
dimensional reduction might not obey the extended rules; 
on the other, the mere presence of subdivergencies changes the simple procedure 
discussed here. In the best of the
worlds, the extended CDR would preserve gauge invariance and supersymmetry, and
not suffer from inconsistencies as the ones in dimensional reduction.

\section*{Acknowledgements}
We thank J.I. Latorre for discussions, and the organizers for a 
pleasant meeting and for their patience. 
This work has been supported by CICYT, under
contract number AEN96-1672 and by Junta de Andaluc\'{\i}a, FQM101.



\begin{thebibliography}{99}

\bibitem{dimreg} G. 't  Hooft and M. Veltman, \np{B44}{72}{189}; 
                 C.G. Bollini and J. Giambiagi, \nc{12 B}{72}{20};
                 J.F. Ashmore, \vj{Nuovo Cim. Lett.}{4}{72}{289}; 
                 G.M. Cicuta and E. Montaldi, 
                 \vj{Nuovo Cim. Lett.}{4}{72}{329}.
\bibitem{dimred} W. Siegel, \pl{B84}{79}{193};
                 D.M. Capper, D.R.T. Jones and P. van Nieuwenhuizen,
                 \np{B167}{80}{479}.
\bibitem{Avdeev83} L.V. Avdeev and A.A. Vladimirov, 
                   \np{B219}{83}{262}.
\bibitem{Siegel2} W. Siegel, \pl{B94}{80}{37}.
\bibitem{Avdeev} L.V. Avdeev, G.A. Chochia and A.A. Vladimirov, 
                 \pl{105B}{81}{272}.
\bibitem{FJL} D.Z. Freedman, K. Johnson and J.I. Latorre, 
              \np{B371}{92}{353}.
\bibitem{QCD} M. P\'erez-Victoria, UG-FT-89/98, hep-th/9808071, to appear in
              {\sl Phys. Lett.} {\bf B}.
\bibitem{g2} F. del Aguila, A. Culatti, R. Mu\~noz Tapia and
                M. P\'erez-Victoria, \np{B504}{97}{532}.
\bibitem{Barna} F. del Aguila, A. Culatti, R. Mu\~noz Tapia and
    M. P\'erez-Victoria, International Workshop on
    Quantum Effects in MSSM, Universitat Aut\`onoma de
    Barcelona, September 1997, hep-ph/9711474.
\bibitem{CDR} F. del Aguila, A. Culatti, R. Mu\~noz Tapia and
    M. P\'erez-Victoria, \pl{B419}{98}{263}; F. del Aguila and M.
    P\'erez-Victoria, 
    \vj{Acta Phys. Polon.}{B28}{97}{2279}.
\bibitem{techniques} F. del Aguila, A. Culatti,
    R. Mun\~oz Tapia and M. P\'erez-Victoria, MIT-CTP-2705,
    UG-FT-86/98, hep-ph/9806451, to appear in {\sl Nucl. Phys.} {\bf B}.   
\bibitem{auto} T. Hahn and M. P\'erez-Victoria, 
    UG-FT-87/98, KA-TP-7-1998, hep-ph/9807565, 
    to appear in {\sl Comp. Phys. Comm.}
\bibitem{Nuria}  G. Dunne and N. Rius, \pl{B293}{92}{367}.
\bibitem{Ferrara} S. Ferrara and E. Remiddi, \pl{B53}{74}{347}.
\bibitem{Mendez}  F. del Aguila, A. M\'endez and F.X. Orteu, 
                  \pl{B145}{84}{70}.
\bibitem{Bellucci}  S. Bellucci, H. Cheng and S. Deser, 
                    \np {B252}{85}{389}.
\bibitem{Wilcox} W. Wilcox, \ap{139}{82}{48}.
\bibitem{Grisaru} M.T. Grisaru and D. Zanon, 
                  \vj{Class. Quant. Grav.}{2}{85}{477}.
\bibitem{Berends} F.A. Berends and R. Gastmans,  \pl{B55}
                  {75}{311}.
\bibitem{Rheinsberg} F. del Aguila and M. P\'erez-Victoria, 
                     \vj{Acta Phys.Polon.}{B29}{98}{2857}.
\bibitem{ABJ} S. Adler, \pr{177}{69}{2426}; 
     J.S. Bell and R. Jackiw, \vj{Nuovo Cimento}{51}{69}{47}.

\end{thebibliography}
\end{document}